\documentstyle[epsfig,prl,twocolumn,aps,floats]{revtex}
\begin{document}
\topmargin = -2.0cm
\overfullrule 0pt
\twocolumn[\hsize\textwidth\columnwidth\hsize\csname
@twocolumnfalse\endcsname
\title
{ New Conditions for a Total Neutrino Conversion in a Medium
}
\author
{ M. V. Chizhov$^1$ and S. T. Petcov$^{2,3}$}
\address
{
$^1$Centre for Space Research and Technologies, Faculty of Physics,
University of Sofia, 1164 Sofia, Bulgaria\\
$^2$Scuola Internazionale Superiore di Studi Avanzati,
Istituto Nazionale di Fizica Nucleare, Sezione di Trieste,\\ 
I-34014, Trieste, Italy\\
$^3$Institute of Nuclear Research and Nuclear Energy, Bulgarian Academy of
Sciences, 1784 Sofia, Bulgaria}
    \maketitle
\vspace{.5cm}
\hfuzz=25pt
\begin{abstract}
A new effect of total neutrino conversion is possible when neutrino
propagates through multi-layer medium of 
nonperiodic constant density layers.
The effect can take place in the oscillations in the Earth 
of the Earth-core-crossing solar and atmospheric neutrinos.
\end{abstract}
\vskip2pc]
\newpage

\medskip

Neutrino oscillations were proposed by B.~Pontecorvo~\cite{Pon} in analogy
with the $K^0-\bar{K}^0$ oscillations (see also \cite{MNS62,Pont67}). 
The pattern of oscillations 
of ultra-relativistic neutrinos 
in vacuum depends in the simplest 
case of two-neutrino mixing on
the vacuum mixing angle $\theta$, 
the ratio of the neutrino 
squared mass difference and the
neutrino energy, $\Delta m^2/E$, 
and the time or distance $t \cong X$ 
traveled by the neutrinos.  
The probability of transition between 
the weak-eigenstate neutrinos 
$\nu_\alpha$ and $\nu_\beta$ ($\alpha \ne \beta = e,\mu,\tau,s$)
in this case, 
\begin{equation}
P_{\alpha\beta}=\sin^2(2\theta)\sin^2\phi~,
\label{Pvac}
\end{equation}
is determined by the angle
$2\theta$ and the phase $\phi=(\Delta m^2/4E) X$.
Therefore, the necessary condition for a 
total neutrino conversion is
$\theta=\pi/4$, while 
the external parameter $X$ 
can, in principle,  be chosen to have
$P_{\alpha\beta}=1$.

   When neutrinos propagate in matter 
an additional phase difference can
arise between the $\nu_\alpha$ and $\nu_\beta$ states 
due to the difference
$V_{\alpha\beta}$ of their effective potentials in matter. In the case
of matter with constant density and chemical composition,
this difference can be accounted for 
through the mixing angle in matter, $\theta_m$, 
and the effective squared mass difference 
$$
\Delta m^2_{eff}=\Delta m^2 \sqrt{
\left(\cos(2\theta)-{2E V_{\alpha\beta} \over \Delta m^2}\right)^2
+\sin^2(2\theta)}~,
$$
in such a way that the transition probability in matter,
\begin{equation}
P_{\alpha\beta}=\sin^2(2\theta_m)\sin^2\phi_m~,
\label{Pmat}
\end{equation}
has the same form as that in vacuum, 
where $\phi_m=(\Delta m^2_{eff}/4E)X$. The  
mixing angle in matter $\theta_m$ 
is given by the familiar resonance-type expression 
$$
\sin^2(2\theta_m)={\sin^2(2\theta) \over \displaystyle
\left(\cos(2\theta)-{2E V_{\alpha\beta} \over \Delta m^2}\right)^2
+\sin^2(2\theta)}~.
$$
It can be maximal, $\theta_m=\pi/4$, even in 
the case of a small vacuum mixing
angle $\theta$, when the resonance condition 
\begin{equation}
{\Delta m^2\over{2E}}\cos(2\theta) - V_{\alpha\beta} = 0,
\label{res}
\end{equation}
\noindent is satisfied, where 
$V_{e\mu} = \sqrt{2}G_F N_e$, $V_{es} = \sqrt{2}G_F (N_e - {1\over 2}N_n)$,
etc., $N_{e,n}$ being the  
electron and neutron number densities of matter.
This is the 
well-known 
MSW effect~\cite{MSW}.

  The neutrino oscillations are a pure quantum 
mechanical phenomenon and it is
interesting to investigate the 
interference effects in 
the neutrino transitions when neutrinos pass 
through a multi-layer medium of (nonperiodic) 
constant density layers. 
In this case the transition 
probability amplitude 
is a sum of products of the 
probability amplitudes of the transitions 
in each layer. In what follows we will show
that due to the constructive interference
between the different amplitudes,
new solutions for a total neutrino conversion 
exist even when the resonance condition 
(\ref{res}) is not fulfilled 
in any of the layers of the medium. 
This phenomenon differs from the 
parametric resonance effect 
possible in a medium 
with periodic density distribution ~\cite{par.res}. 
In order for the new effect 
to take place, the passage of neutrinos 
through two layers of different constant
density is sufficient. 
Moreover, the strong resonance-like 
enhancement of the 
probabilities of the 
$\nu_2 \rightarrow \nu_{e}$,
$\nu_{\mu} \rightarrow \nu_{e}$
($\nu_e \rightarrow \nu_{\mu(\tau)}$),
$\nu_e \rightarrow \nu_{s}$, etc. 
transitions in the Earth of the 
Earth-core crossing solar and atmospheric 
neutrinos (see, e.g., \cite{PastSun,Art2,PastAtmo} 
and the references quoted therein),
is due to this effect. 
It was shown in \cite{SP1} that
the standard interpretation of this
enhancement (see, e.g., \cite{PastSun,PastAtmo}) 
as being due to the MSW effect in the Earth core
is incorrect and it was suggested 
that the enhancement  
is generated by a new type of resonance - 
``neutrino oscillation length resonance'' (NOLR). 
At small mixing angles and in the case of the
transitions $\nu_2 \rightarrow \nu_{e}$ and
$\nu_{\mu} \rightarrow \nu_{e}$, for instance, 
the values of the parameters at which
the maximal neutrino conversion takes place 
for neutrinos traversing the Earth core 
are rather close to the values of the parameters 
at which the NOLR can occur.
In all transitions, 
however, only the maximal neutrino 
conversion mechanism is operative 
for the Earth-core-crossing neutrinos.

   Consider the simple case of media consisting of
two layers having constant but
different densities and chemical compositions.
Then the mixing angles in matter for 
the first and the second layer,
$\theta'_m$ and $\theta''_m$, are
constant. We shall assume without loss of generality that
\begin{equation}
 0 \leq  V'_{\alpha \beta} < V''_{\alpha \beta}, 
\end{equation}
where $V'_{\alpha \beta}$ and $V''_{\alpha \beta}$ 
are the effective
potential differences in the first and the second layer, respectively. 
Our analysis will apply 
to the case when the first layer is vacuum
$V'_{\alpha \beta}=0$ as well, 
because the neutrino oscillations in vacuum
and in constant-density medium are 
described by the same equations (see
(\ref{Pvac}) and (\ref{Pmat})). 
The probability amplitude of the transition 
$\nu_\alpha \to \nu_\beta$ 
after neutrinos have crossed the two layers 
represents a sum of two terms:
\begin{equation}
A_{\alpha\beta}=A'_{\alpha\alpha}A''_{\alpha\beta}+
A'_{\alpha\beta}A''_{\beta\beta},
\label{amplitude}
\end{equation}
where
\begin{eqnarray}
A'_{\alpha \alpha } &=& \cos\phi'_m+i\cos(2\theta_m')\sin\phi'_m,
\nonumber
\\
A''_{\alpha\beta} &=& -i\sin(2\theta_m'')\sin\phi''_m, 
\nonumber
\\
A'_{\alpha\beta} &=& -i\sin(2\theta_m')\sin\phi'_m,
\nonumber
\\
A''_{\beta \beta} &=& \cos\phi''_m-i\cos(2\theta_m'')\sin\phi''_m.  
\end{eqnarray}
Here $A'_{\alpha \alpha}$, $A'_{\alpha \beta}$ 
and  $A''_{\alpha\beta}$, $A''_{\beta \beta}$
are the probability amplitudes describing the 
oscillations in the first and in the second layer, 
respectively,
$\phi'_m=(\Delta m'^{~2}_{eff}/4E)X'$ and $\phi''_m=(\Delta
m''^{~2}_{eff}/4E)X''$ are half of the phase differences 
accumulated by the neutrino 
energy eigenstates 
in the first and in the second layer, and
$X'$ and $X''$ are the lengths of
paths of the neutrinos in the 
corresponding layers. The transition
probability
\begin{eqnarray}
P_{\alpha\beta}=|A_{\alpha\beta}|^2&=&
|A'_{\alpha\alpha}|^2|A''_{\alpha\beta}|^2+
|A'_{\alpha\beta}|^2|A''_{\beta\beta}|^2
\nonumber
\\
&+&2{\rm Re}\left((A'_{\alpha\alpha}A''_{\alpha\beta})^*
A'_{\alpha\beta}A''_{\beta\beta}\right)
\label{P2}
\end{eqnarray}
is a sum of two products of the probabilities 
of the neutrino oscillations
in the different layers and of the interference 
term. The latter can play
a crucial role in the enhancement 
of the neutrino transitions.

  For monoenergetic neutrinos 
the path lengths $X'$ and $X''$ 
can be considered as independent 
variables of the system, as long as the mixing angles 
in matter $\theta'_m$ and $\theta''_m$ are fixed. 
Varying these variables we get a 
system of equations for the local maxima of $P_{\alpha\beta}$, 
\begin{equation}
\displaystyle{{\rm d}P_{\alpha\beta} \over {\rm d}\phi'_m}=0, ~~~~~
\displaystyle{{\rm d}P_{\alpha\beta} \over {\rm d}\phi''_m}=0,
\label{extr}
\end{equation}
provided the supplementary conditions 
on the values of the second derivatives
of $P_{\alpha\beta}$ at the points where 
equations (8) hold,
\begin{equation}
\displaystyle{{\rm d}^2 P_{\alpha\beta} \over {\rm d}\phi'^{~2}_m}<0,
~~~~~
\displaystyle{{\rm d}^2 P_{\alpha\beta} \over {\rm d}\phi'^{~2}_m}
{{\rm d}^2 P_{\alpha\beta} \over {\rm d}\phi''^{~2}_m}-
\left(
{{\rm d}^2 P_{\alpha\beta} \over {\rm d}\phi'_m {\rm d}\phi''_m}
\right)^2 > 0,
\end{equation}
are fulfilled. This system has four types of different solutions 
determining the local and absolute maxima of 
the transition probability. 
The first one
\begin{equation}
solution~I:\left\{ \begin{array}{l}
\cos\phi'_m=0,~or~ 2\phi'_m = \pi (2k' + 1),\\
\sin\phi''_m=0,~or~ 2\phi''_m = 2\pi k'',
\end{array} \right.
\label{sI}
\end{equation}
where $k',k''= 0,1,...$, is valid in the region $I$ (Fig.~1)
\begin{equation}
region~I:~ \cos(2\theta'_m)\le0
\label{rI}
\end{equation}
and corresponds to a neutrino conversion 
in the first layer only: 
$A''_{\alpha\beta}=0$.
In this case the interference term in (\ref{P2}) is zero and the 
transition probability in the two layers 
is defined by the transition
probability in the first layer
\begin{equation}
type~I:~ P^{max}_{\alpha \beta} = \sin^2(2\theta'_m).
\end{equation}
A total neutrino conversion, $P_{\alpha\beta}=1$, occurs when the
MSW resonance condition in the first layer,  
\begin{equation}
\cos(2\theta'_m)=0,
\label{cI}
\end{equation}
is satisfied. 
Thus, the absolute maxima in region $I$ can be ascribed 
to the MSW effect in the first layer. 

  The second solution,
\begin{equation}
solution~II:\left\{ \begin{array}{l}
\sin\phi'_m=0,~or~ 2\phi'_m = 2\pi k',\\
\cos\phi''_m=0,~or~ 2\phi''_m = \pi (2k'' + 1),
\end{array} \right.
\label{sII}
\end{equation}
$k',k''=0,1,...$, is realized in region $II$ (Fig.~1),
\begin{equation}
region~II:~ \cos(2\theta''_m)\ge 0,
\label{rII}
\end{equation}
and is completely analogous to the previous one, with the two layers
interchanged. The phase requirements (\ref{sII}) reduce again the
expression (\ref{P2}) to only one term
\begin{equation}
type~II:~ P^{max}_{\alpha \beta} = \sin^2(2\theta''_m),
\end{equation}
with no contribution from the interference term as
$A'_{\alpha\beta}=0$. Thus, a total neutrino 
conversion in this case can be 
ascribed to  the MSW effect in the second layer. 

   We get a very different mechanism of enhancement of the neutrino
transition probability in the region
\begin{equation}
\left\{ \begin{array}{l}
\cos 2\theta_m' > 0, \\
\cos 2\theta_m'' < 0, \\
\end{array} \right.
\label{intermediate}
\end{equation}
located between the regions $I$ and $II$. The maxima of
$P_{\alpha\beta}$ in this region are caused by maximal contribution of the
interference term in equation (\ref{P2}) for $P_{\alpha\beta}$,
taking place when the two
vectors ${\bf z}_1=({\rm Re}(A'_{\alpha\alpha}A''_{\alpha\beta})$, 
${\rm Im}(A'_{\alpha\alpha}A''_{\alpha\beta}))$ and 
${\bf z}_2=({\rm Re}(A'_{\alpha\beta}A''_{\beta\beta})$, 
${\rm Im}(A'_{\alpha\beta}A''_{\beta\beta}))$ 
in the complex plane are
collinear and point in the same direction.

      Indeed, the third solution,
\begin{equation}
solution~III:\left\{ \begin{array}{l}
\cos\phi'_m=0,~or~ 2\phi'_m = \pi (2k' + 1)\\
\cos\phi''_m=0,~or~ 2\phi''_m = \pi (2k + 1)
\end{array} \right.
\label{sIII}
\end{equation}
\noindent where $k',k=0,1,...$, leads to local maxima,
\begin{figure}[htb]
\epsfig{file=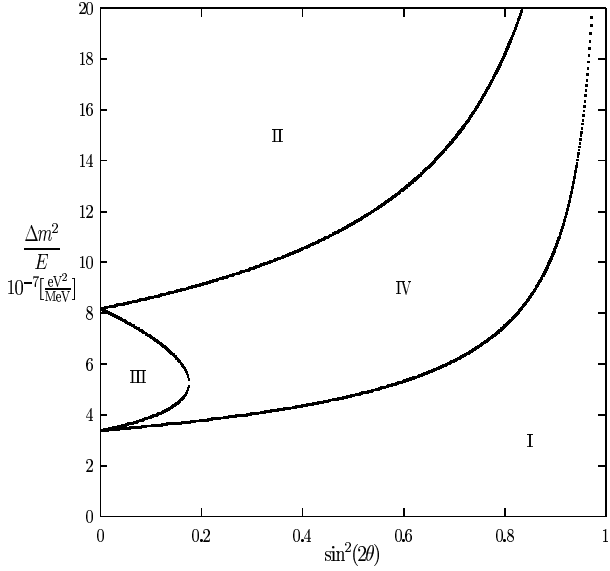,height=7.5cm,width=8.0cm}
\vglue 0.4cm 
\caption{The regions of
the four different solutions for the maxima of the
transition probability $P_{e\mu} = P_{\mu e}$
in a two-layer medium. The two different layers
correspond to the mantle and the core of the Earth.}
\label{fig1} 
\end{figure}
\begin{equation}
type~III:~ P^{max}_{\alpha \beta} = \sin^2 (2\theta'_m-2\theta''_m),
\end{equation}
if the oscillation parameters belong to the 
finite region $III$ (Fig.~1),
\begin{equation}
region~III:~ \cos(2\theta'_m-2\theta''_m)\le 0.
\label{rIII}
\end{equation}
This maximum demonstrates the 
new effect of enhancement of the 
probability $P_{\alpha \beta}$, 
when the interference between the 
two terms in the r. h.
side of
(\ref{amplitude}) is maximal and constructive 
\footnote{The two-layer solutions (\ref{sIII})
are the analogs of the solutions
corresponding to 
the neutrino oscillation length
resonance \cite{SP1} 
in the case of the transitions
in the Earth  
of the Earth-core-crossing neutrinos.}. 
However, solution III does
not provide an absolute maximum 
of $P_{\alpha \beta}$, except on the curve
\begin{equation}
\cos(2\theta'_m - 2\theta''_m)=0.
\label{cIII}
\end{equation} 
 Only the solutions of the fourth type,
\begin{equation}
solution~IV:\left\{ \begin{array}{l}
\tan\phi'_m=\pm\sqrt{{\displaystyle -\cos(2\theta_m'')\over
\displaystyle\cos(2\theta_m')\cos(2\theta_m'' - 2\theta_m')}}~ \\
\tan\phi''_m=\pm\sqrt{{\displaystyle -\cos(2\theta_m')\over
\displaystyle\cos(2\theta_m'')\cos(2\theta_m'' - 2\theta_m')}}~
\end{array} \right.
\label{sIV}
\end{equation}
where the signs are correlated,
provide an absolute maximum 
of $P_{\alpha \beta}$,   
\begin{equation}
type~IV:~ P^{max}_{\alpha \beta} = 1,
\end{equation}
in the region $IV$ (Fig.~1),
\begin{equation}
region~IV:~\left\{ \begin{array}{l}
\cos 2\theta_m' \ge 0, \\
\cos 2\theta_m'' \le 0, \\
\cos(2\theta_m''-2\theta_m') \ge 0.
\end{array} \right.
\label{rIV}
\end{equation}
Note that these solutions do not imply
any additional constraints on the mixing angles in matter
of the type of the
resonance conditions (\ref{cI}) or (\ref{cIII}).
In contrast to the case of the MSW
effect, in which a total neutrino conversion 
can be realized only on the
curve (\ref{res}) in the 
$\sin^2(2\theta)-\Delta m^2/E$  plane of 
parameters, the new effect is possible in the whole 
2-dimensional region $IV$.

      When the widths of the layers $X'$ and $X''$ 
are fixed, the neutrino energy $E$ is the only 
physical parameter which can be varied.
The extrema condition reads in this case:
\begin{eqnarray}
{{\rm d}P_{\alpha\beta} \over {\rm d}E} &=& 
{{\rm d}P_{\alpha\beta} \over {\rm d}\phi'_m}
{{\rm d}\phi'_m \over {\rm d}E}
 + {4V'_{\alpha\beta} \over \Delta m'^{~2}_{eff}}
\sin(2\theta'_m)
\nonumber
\\
&\times& \cos(2\theta'_m-2\theta''_m)\sin\phi'_m\cos\phi''_m
\nonumber
\\
&\times& \left\{
\left[{\cos(2\theta'_m)\over
\cos(2\theta''_m)\cos(2\theta'_m-2\theta''_m)} + \tan^2\phi''_m
\right]
\nonumber \right.
\\
&\times& \sin(2\theta'_m)\cos(2\theta''_m)\sin\phi'_m\cos\phi''_m
\nonumber
\\
&+& \cos(2\theta'_m)\sin(2\theta''_m)\cos\phi'_m\sin\phi''_m
\nonumber
\\
&\times& \left.
\left[{1 \over\cos(2\theta'_m-2\theta''_m)}-\tan\phi'_m \tan\phi''_m
\right] \right\}
\nonumber
\\
&+&(' \leftrightarrow '')=0.
\end{eqnarray}
It is clear, that, in general, 
the solutions of the system (\ref{extr})
do not correspond to extrema in 
the variable $E$. Only the solutions 
(\ref{sIV}) for the absolute maxima, 
which correspond to a total neutrino conversion, give the
absolute maxima in any variable. The solutions (\ref{sI}),
(\ref{sII}), and (\ref{sIII}) for
the local maxima with $P_{\alpha\beta}<1$ no longer
correspond to extrema in the variable $E$. New solutions are
possible, which do not coincide 
with the solutions corresponding to phases
equal to multiples of $\pi$. 
As can be shown, the same conclusion 
is valid  also if  
the width of one of the layers is 
determined by the width of
the second layer. The indicated possibilities
are realized, e.g., for neutrinos born in the
Earth central region.

  In the case of the solar and atmospheric 
neutrinos passing through the
Earth core on the way to the detectors, 
the Earth density distribution 
can be approximated by 
a two-layer density profile, 
core - mantle, and neutrinos cross three layers:
mantle - core - mantle. For the calculation 
of the probabilities of interest
the two-layer model of the Earth 
provides a very good approximation 
to the more complicated density distributions
predicted by the existing models of the Earth
(see, e.g., \cite{SP1,MP98:2layers,KP3nu88}).
Because of the spherical symmetry of the Earth, 
the lengths $X'$ and $X''$ are not 
independent variables. 
However, as is shown in ~\cite{ChPet99B}, the 
analogs of the solutions 
$IV$, corresponding to 
\begin{figure}[htb]
\epsfig{file=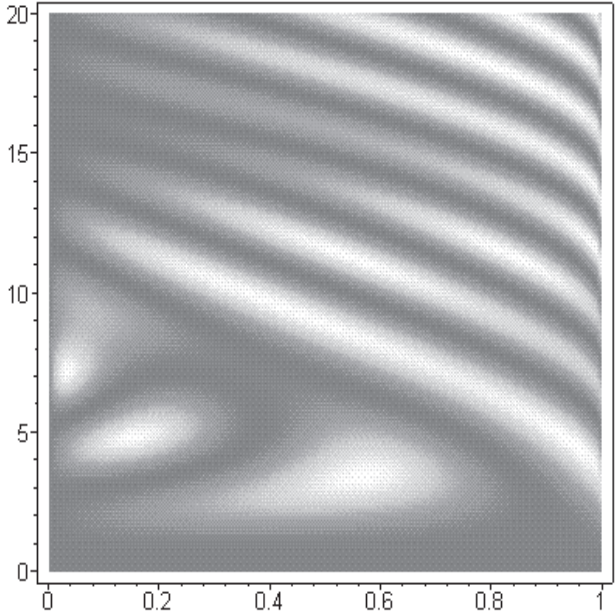,height=8.5cm,width=8.4cm}
\vglue 0.4cm 
\caption{The probability 
$P_{e\mu} = P_{\mu e}$ for the Earth center crossing
neutrinos, as a function of $\sin^22\theta$ 
(horizontal axis) and $\Delta m^2/E~[10^{-7}~{\rm eV^2/MeV}$]
(vertical axis). The 
greyscales correspond to
different values of $P_{e\mu}$: 
0.0 - 0.10 is represented by black, etc., while 
the white represents values 
in the interval 0.9 - 1.0. 
The points of total neutrino 
conversion (in the white 
regions), $P_{e\mu} = 1$,
correspond to solutions $IV$ 
for the Earth-core-crossing 
neutrinos, eq. (\ref{Earth}).}
\label{fig2} 
\end{figure}
\noindent
a total neutrino conversion, $P_{\alpha\beta}=1$, 
exist for the Earth
density profile as well. 
For the
$\nu_{\mu (e)} \rightarrow \nu_{e (\mu ;\tau)}$,
$\nu_e \rightarrow \nu_{s}$ and
$\bar{\nu}_{\mu} \rightarrow \bar{\nu}_{s}$
transitions ($\Delta m^2~\cos 2\theta > 0$) they read:
\begin{eqnarray} 
\tan\phi'_{m}&=&\pm\sqrt{{\displaystyle -\cos 2\theta_m''\over
\displaystyle\cos(2\theta_m''- 4\theta_m')}},
\nonumber
\\
\tan\phi''_{m}&=&\pm{\displaystyle \cos 2\theta_m'\over \sqrt{
\displaystyle-\cos(2\theta_m'')\cos(2\theta_m''- 4\theta_m')}},
\label{Earth}
\end{eqnarray}
where the signs are correlated;
the solutions for the
$\nu_2 \rightarrow \nu_{e}$ transitions
can formally be obtained from eq. (\ref{Earth})
by replacing $2\theta_m''$ and $2\theta_m'$ in the
expressions for $\phi'_{m}$ and $\phi''_{m}$
with $2\theta_m'' - \theta$ and $2\theta_m' - \theta$.
Moreover, these solutions 
are responsible for the 
strong resonance-like enhancement
of the $\nu_2 \rightarrow \nu_{e}$,
$\nu_{\mu} \rightarrow \nu_{e}$,
$\nu_e \rightarrow \nu_{\mu(\tau)}$,
$\nu_e \rightarrow \nu_{s}$, etc. 
transitions in the Earth of the 
Earth-core crossing solar and atmospheric 
neutrinos, discussed in \cite{SP1} 
(see also  \cite{s5398}).
In \cite{SP1} this enhancement was 
interpreted to be due to the
neutrino oscillation length resonance 
(NOLR) -
the analog of the solution
(\ref{sIII}) in the case of the 
Earth-core-crossing neutrinos.
On the boder line between the 
regions where the NOLR and the maximal
conversion solution (\ref{Earth})
can be realized, the two solutions coincide
and we have $P^{max}_{\alpha \beta} = 1$
at the points where they hold.
However, the NOLR does not describe
the local maxima, 
$P^{max}_{\alpha \beta} < 1$,
in the variable $\Delta m^2/E$  
at fixed  $\sin^22\theta$ 
and/or fixed $h$ (for further details
see \cite{ChPet99B}).
In all transitions of interest only 
the maximal neutrino 
conversion mechanism is operative 
for the Earth-core-crossing neutrinos,
both at small and large 
values of  $\sin^22\theta$ (Fig. 2).
 
\vskip 0.3cm
\noindent {\bf Acknowledgements.} This work 
was supported in part by the Italian MURST
under the program ``Fisica Teorica delle
Interazioni Fondamentali''; the work of S.T.P.
was partially supported also by Grant PH-510 from the
Bulgarian Science Foundation.


\baselineskip 18pt
\begin{thebibliography}{99}
\bibitem{Pon} B. Pontecorvo, Zh. Eksp. Teor. Fiz. {\bf 33}, 549 (1957); 
{\it ibid.} {\bf 34}, 247 (1958).

\bibitem{MNS62} Z. Maki, M. Nakagawa and S. Sakata,
Prog. Theor. Phys. {\bf 28}, 870 (1962).

\bibitem{Pont67} B. Pontecorvo, Zh. Eksp. Teor. Fiz. {\bf 53}, 1717 (1967);
V. Gibov and B. Pontecorvo, Phys. Lett. {\bf B28}, 493 (1969).
S.M. Bilenky and B. Pontecorvo, Phys. Rep. {\bf 41}, 225 (1978).

\bibitem{MSW} S.P. Mikheyev and A.Yu. Smirnov, Sov. J. Nucl. 
Phys. {\bf 42}, 913 (1985); 
L. Wolfenstein, Phys. Rev. {\bf D17},2369 (1978).

\bibitem{par.res}  V.K. Ermilova et al., Short Notices of
the Lebedev Institute {\bf 5}, 26 (1986); 
E. Akhmedov, Yad. Fiz. {\bf 47}, 475
(1988); P. Krastev and A.Yu. Smirnov, 
Phys. Lett. {\bf B226}, 341 (1989).

\bibitem{PastSun} M. Cribier et al.,
Phys. Lett. {\bf{B182}}, 89 (1986);
A.J. Baltz and J. Weneser,
Phys. Rev. {\bf D35}, 528 (1987), D{\bf 50}, 5971 (1994);
J.M. LoSecco, Phys. Rev.  {\bf D47}, 2032 (1993);
 J.M. Gelb, W. Kwong and S.P. Rosen, Phys. Rev. Lett. {\bf 78}, 2296 (1997).

\bibitem{Art2} M. Maris and S.T. Petcov, Phys. Rev. D {\bf 56}, 7444 (1997).

\bibitem{PastAtmo} S.P. Mikheyev and A.Yu. Smirnov, in
{\it Massive Neutrinos
in Astrophysics and in Particle Physics},
Proceedings of the Moriond Workshop 
1986, Tignes, France,
(eds. O. Fackler and J. Tran Thanh Van,
Editions Fronti{\`e}res, Gif-sur-Yvette, 1986), p. 355;
E.D. Carlson, Phys. Rev. {\bf D34}, 1454 (1986);
A. Dar et al., Phys. Rev. {\bf D35}, 3607 (1987).

\bibitem{SP1} S.T. Petcov, Phys. Lett. {\bf B434}, 321 (1998),
(E) {\bf B444}, 584 (1998).

\bibitem{MP98:2layers} M. Maris and S.T. Petcov, to be published.

\bibitem{KP3nu88} P.I. Krastev and S.T. Petcov, Phys. Lett.
{\bf B205}, 84 (1988).

\bibitem{ChPet99B} M. V. Chizhov and  S. T. Petcov, Report SISSA 28/99/EP,
March 1999 (hep-ph/9903424).

\bibitem{s5398} M. Chizhov,
M. Maris and S.T. Petcov, Report SISSA  53/98/EP, 31 July 1998
(hep-ph/9810501).

\end{thebibliography}
\end{document}